\documentclass{article}

\title{A Derivation of the Quantized Electromagnetic Field Using Complex Dirac Delta Functions}
\author{Robert Ducharme}

\begin{document}

\maketitle

\centerline{2112 Oakmeadow Pl., Bedford, TX 76021}
\centerline{E-mail: robertjducharme66@gmail.com}

\begin{abstract} 
It is shown a complex function $\Phi$ defined to be the product of a real Gaussian function and a complex Dirac delta function satisfies the Cauchy-Riemann equations. It is also shown these harmonic $\Phi$-functions can be included in the solution of the classical electromagnetic field equations to generate the quantum field as a many-particle solution such that the $\Phi$-functions represent the particle states. Creation and destruction operators are defined as usual to add or subtract photons from the particle states. The orbital angular momentum of the $\Phi$-states is interpreted as spin since it emerges from a point source that must be circularly polarized as a requirement of the gauge condition.
\end{abstract}

\section{Introduction}
Electromagnetic radiation has both wave and particle properties. The classical field equations are known to describe the wave properties. The particle properties are conventionally introduced through a field quantization procedure \cite{IZ} that decomposes the field into an electromagnetic field operator and particle states. The purpose of this paper is to show solutions to the classical field equations can be constructed to include particle states defined in terms of complex point functions. It is also to demonstrate the inclusion of these particle states is equivalent to field quantization.

Complex point functions take the form $\phi_{\beta} = \psi(|z_{\beta}|)\delta(z_{\beta}^*-\xi_{\beta}^*)$ where $\psi$ is a Gaussian function, $\delta$ is a Dirac delta function and $\beta = \pm 1$. In this $z_\beta = x_1 + \imath \beta x_2$ is a complex coordinate and $\xi_{\beta} = \xi_a + \imath \beta \xi_b$ is an arbitrary complex constant. Complex point functions satisfy the Cauchy-Riemann (C-R) equations and are therefore harmonic functions. They can be included in solutions of Maxwell's equations as a result of being harmonic functions and satisfy the Lorenz gauge condition since it reduces to the form of the C-R equations for the case of circular polarization. The modulus $|\phi_{\beta}|$ is restricted to a single point $(x_1 = \xi_a, x_2 = \xi_b)$. 

It is well known that Dirac delta functions of position coordinates can be used to represent constraint spaces. The theory of real constraint spaces has been extensively developed in the context of constraint mechanics \cite{AK, RST, CA, WC}. The properties of complex point functions will be developed in section 2 of this paper. It will be argued that the role of the complex delta function \cite{RM,JGM} in complex point functions is to constrain the real Gaussian function $\psi$ to one of two possible domains denoted by $\mathbf{C}_{\beta}$. The $\phi_\beta$-functions in these domains are related to each other through the parity transformation $(x_1 \rightarrow x_1, x_2 \rightarrow -x_2)$. Functions in the two $\mathbf{C}_{\beta}$ domains are therefore distinguishable as the mirror images of each other.

In using $\phi_\beta$-functions to represent photons it is important to be mindful that photons can be identical. Let $\phi_\beta^j$ denote an element in a set of component functions for $n$ identical photons. The set of $n$ photons can then be represented using a symmetrical many-particle wavefunction $\Phi_\beta^n$ constructed in terms of the $\phi_\beta^j$ component functions.

In section 3, Maxwell's equations are solved alongside the Lorenz gauge condition for classical electromagnetic waves subject to the C-R equations. The role of the C-R equations is to confine the electromagnetic field to the two $\mathbf{C}_{\beta}$ domains where the $\Phi_{\beta}^n$-functions are defined. The $\Phi_{\beta}^n$-functions emerge as a component in each circularly polarized Fourier node $k$. It will be shown the gauge condition correlates the right and left circular polarization states to different values of $\beta$.

Creation and annihilation operators can be defined to add and subtract photons from $\Phi_\beta^n$-functions. It is this assumption that enables the solution to the classical electromagnetic field equations to be expressed in the form of a quantum field consisting of a field operator containing creation and annihilation operators as well as particle states expressed in terms of the $\Phi_\beta^n$-functions. The index n in each $\Phi_{\beta}^n$-functions is thus interpreted as the occupancy number of indistinguishable particles having the same wave vector and polarization state.

One consequence of obtaining the quantized form of the electromagnetic field from the classical field equations is that the concept of a many-particle wavefunction is retained. In section 4, it is shown quantum mechanical operators can be applied to this wavefunction to determine the properties of the quantum field. The method gives the same results as applying conventional quantum field theory (QFT) operators direct to the particle states. 

In the classical theory of the electromagnetic field, the total angular momentum of the field is the sum of contributions from orbital angular momentum (OAM) and spin angular momentum (SAM). It has been stated the role of the complex point functions in the electromagnetic field is to represent the particle states. It is shown the angular momentum of $\Phi_{\beta}^{n}$-functions can be calculated using the same quantum mechanical angular momentum operator used to calculate OAM in real space. It will, however, be interpreted as SAM since it is a local property of a point and correlates to polarization. It is in fact spin that distinguishes the two $\mathbf{C_\beta}$ domains from each other. 

\section{Complex Point Functions}
A point $(x_1,x_2)$ in the real plane can be expressed in the complex plane $\mathbf{C}$ using the coordinates
\begin{equation} \label{eq: conftrans1}
z_\beta = x_1 + \imath \beta x_2
\end{equation}
where $z_{+1}$ and $z_{-1}$ denote complex conjugate quantities.  The inverse of these relationships can be written
\begin{equation} \label{eq: inv_ct}
x_1 = \frac{z_{+1} + z_{-1}}{2}, \quad x_2 = \frac{z_{+1} - z_{-1}}{2\imath}
\end{equation}
Eqs. (\ref{eq: inv_ct}) can be used alongside the chain rule for partial differentiation to derive the Wurtinger derivatives \cite{GR}
\begin{equation} \label{eq: complexDiff1}
\frac{\partial}{\partial z_\beta}  
= \frac{1}{2} \left( \frac{\partial}{\partial x_1}
- \imath \beta \frac{\partial}{\partial x_2} \right),
\end{equation}
These results assume
\begin{equation} \label{eq: complexDiff5}
\frac{\partial z_{+1}}{\partial z_{-1}}=\frac{\partial z_{-1}}{\partial z_{+1}} = 0
\end{equation}
enabling $z_{+1}$ and $z_{-1}$ to be treated as independent coordinates. 

In order for complex differentiation to be meaningful, the result cannot depend on the direction of differentiation in the complex plane. This requirement leads to the C-R equations
\begin{equation} \label{eq: cr1}
\frac{\partial f}{\partial z^*} = \frac{\partial f}{\partial x_1} + \imath \frac{\partial f}{\partial x_2} = 0
\end{equation}
for $z^*=z_{-1}$; and the complex conjugate form
\begin{equation} \label{eq: cr2}
\frac{\partial f}{\partial z} = \frac{\partial f}{\partial x_1} - \imath \frac{\partial f}{\partial x_2} = 0
\end{equation}
for $z=z_{+1}$. It is clear a function must be of the form $f(z_\beta)$ to satisfy one of these conditions. The function $f(z_\beta)$ is complex differentiable at a point $z_p$ if it is analytic in the neighborhood of a $z_p$. It is also holomorphic on the domain D if it is complex differentiable at every point in D.

The argument above suggests a real function $\psi(|z_\beta|)$ cannot be complex differentiable on the domain $z_\beta,z_\beta^* \in \mathbf{C}$ owing to the dependence of the function on $z_\beta^*$. Now consider the function
\begin{equation} \label{eq: pf1}
\phi_{\beta} = \psi\left(|z_\beta|\right) \delta\left(z_\beta^*-\xi_\beta^* \right)
\end{equation}
where the delta function has a complex argument and $\xi_\beta = \xi_a + \imath \beta \xi_b$ is a complex constant. It can be seen $\phi_{\beta}$ vanishes except on the domain
\begin{equation} \label{eq: dom1}
\mathbf{C}_{\beta} = \{ z_{\beta}, z_{\beta}^* \in \mathbf{C} | z_{\beta}^* = \xi_\beta^* \}
\end{equation}
where $\psi(|z_\beta|)$ satisfies the C-R equations owing to the fact $z^*$ is held constant. It is of interest next to investigate if $\phi_{\beta}$-functions themselves satisfy the C-R equations.

The bulk of the mathematical literature on the Dirac delta function assumes a real argument. It is helpful therefore to start from the definition
\begin{equation} \label{eq: ddef1}
\int_{-z_0}^{+z_0} \delta(z-\xi_\beta) dz = \int_{-\tau_0}^{+\tau_0} \delta(\tau) d\tau = 1
\end{equation}
where 
\begin{equation} \label{eq: subs}
x_1 = \alpha_1 \tau + \xi_a, \quad \beta x_2 = \alpha_2 \tau + \beta \xi_b, \quad z_0 = (\alpha_1 + \imath \alpha_2) \tau_0 + \xi_\beta, 
\end{equation}
$\tau$ is a real variable and $\alpha_1, \alpha_2, \tau_0$ are real constants $(\tau_0 > 0)$. This substitution shows an integral over a complex delta function is equivalent to an integral over a real delta function. It follows
\begin{equation} \label{eq: ddef2}
\int_{-z_0}^{+z_0} \psi(|z_\beta|)\delta(z_\beta-\xi_\beta) dz_\beta = \int_{-\tau_0}^{+\tau_0} \psi(|\tau+\sigma_\beta|)\delta(\tau) d\tau = \psi(|\xi_\beta|)
\end{equation}
where $\xi_\beta = (\alpha_1 + \imath \alpha_2) \sigma_\beta$ having made the additional assumption $\alpha_1^2 + \alpha_2^2 = 1$ such that $|\xi_\beta| = |\sigma_\beta|$.

The expression
\begin{equation} \label{eq: contour1}
\int_{-z_0^*}^{+z_0^*} \frac{\partial \phi}{\partial z^*} dz^* = \int_{-\tau_0}^{+\tau_0} \psi^\prime (|\tau+\sigma_\beta|) \delta(\tau)d\tau + \int_{-\tau_0}^{+\tau_0}  \psi(|\tau+\sigma_\beta|) \delta^\prime (\tau)d\tau 
\end{equation}
can be evaluated using the standard integrals
\begin{equation} \label{eq: standard1}
\int_{-\tau_0}^{+\tau_0} \psi^\prime (|\tau+\sigma_\beta|) \delta(\tau)d\tau = +\psi^\prime(|\sigma_\beta|)
\end{equation}
\begin{equation} \label{eq: standard2}
\int_{-\tau_0}^{+\tau_0}  \psi(|\tau+\sigma_\beta|) \delta^\prime (\tau)d\tau = -\psi^\prime(|\sigma_\beta|)
\end{equation}
to give
\begin{equation} \label{eq: contour2}
\int_{-z_0^*}^{+z_0^*} \frac{\partial \phi}{\partial z^*} dz^* = 0
\end{equation}
Eq. (\ref{eq: contour2}) vanishes if the integrand vanishes giving
\begin{equation} \label{eq: diff_pf}
\frac{\partial \phi_{\beta}}{\partial z_\beta^*} = 0
\end{equation}
This result confirms $\phi_{\beta}$ satisfies the C-R eqs. (\ref{eq: cr1}) for $\beta=+1$ and eqs. (\ref{eq: cr2}) for $\beta=-1$. 

For applications to the electromagnetic field, it will be convenient to use complex point functions of the specific form
\begin{equation} \label{eq: cpf_qft}
\phi_{\beta}^j = \exp \left[ - \frac{ \kappa^2}{2}(|\zeta_\beta^j|^2-1)  \right]\delta(\zeta_\beta^{j*}-1)
\end{equation}
where 
\begin{equation} \label{eq: zeta}
\zeta_\beta^j = \frac{z_\beta}{\xi_\beta^j} = \frac{x_1 + \imath \beta x_2}{\xi_a^j + \imath \beta \xi_b^j}
\end{equation}
$0 < j \leq n$ is a particle index for a photon belonging to a set of $n$ identical photons and $\kappa$ is a constant. Eq. (\ref{eq: cpf_qft}) satisfies the normalization condition
\begin{equation} \label{eq: normalize}
\int_{-z_0}^{+z_0} \int_{-z_0^*}^{+z_0^*} \phi_{\beta}^{j*}\phi_{\beta}^j dz dz^* = 1
\end{equation}
The total symmetric \cite{WG} wavefunction for n identical photons is
\begin{equation} \label{eq: total_symmetric}
\Phi_{\beta}^n = \frac{1}{\sqrt{n!}}\sum_{P} P(\phi_{\beta}^1 \phi_{\beta}^2 .... \phi_{\beta}^n)
\end{equation}
where the sum is over the different permutations of the positions of the particles.

Eqs. (\ref{eq: diff_pf}) and (\ref{eq: total_symmetric}) show 
\begin{equation} \label{eq: cr3} 
\frac{\partial \Phi_{\beta}^n}{\partial z_\beta^*} = 0
\end{equation}
and therefore
\begin{equation} \label{eq: cr4} 
\frac{\partial^2 \Phi_{\beta}^n}{\partial z_\beta \partial z_\beta^*} = \frac{\partial^2\Phi_{\beta}^{n}}{\partial x_1^2} + \frac{\partial^2\Phi_{\beta}^{n}}{\partial x_2^2} = 0
\end{equation}
This result gives an indication of how the particle state function $\Phi_{\beta}^n$ might be included in the solution to Maxwell's equations since Maxwell's equations contain a Laplacian operator.

\section{The Quantized Electromagnetic Field}
Electromagnetic radiation can be represented using a 4-potential $A_\mu (x_\nu)$  where $\mu, \nu = 0,1,2,3$ and $x_\mu $ is position in Minkowski 4-space. The classical field equations for $A_\mu$  consist of Maxwell's equations 
\begin{equation} \label{eq: maxwell} 
\frac{\partial^2 A_\mu}{\partial x_1^2} + \frac{\partial^2 A_\mu}{\partial x_2^2} + \frac{\partial^2 A_\mu}{\partial x_3^2} - \frac{1}{c^2}\frac{\partial^2 A_\mu}{\partial t^2} = 0
\end{equation}
(having put $t=x_0$) and the Lorenz gauge condition
\begin{equation} \label{eq: lorenz_gauge} 
\frac{\partial A_1}{\partial x_1} + \frac{\partial A_2}{\partial x_2} + \frac{\partial A_3}{\partial x_3} - \frac{\partial A_0}{\partial t} = 0
\end{equation}
where $c$ is the velocity of light. 

The general solution to Maxwell's equations (\ref{eq: maxwell}) can be written in the form of the Fourier expansion
\begin{equation} \label{eq: em_waves} 
A_{\mu}(x_\nu) = \int \frac{d^3k}{(2\pi)^{3/2} \sqrt{2\omega_k}} \sum_{\beta} [a_{\beta}(\vec{k})\epsilon_{\mu \beta}^{k} e^{\imath(\vec{k}.\vec{x}-\omega t)} + a_{\beta }^{*}(\vec{k})\epsilon_{\mu \beta}^{k*} e^{-\imath(\vec{k}.\vec{x}-\omega t)}]
\end{equation}
In this, $\epsilon_{\mu\beta}^k$ is the polarization vector, $a_{\beta}(\vec{k})$ the amplitude of each wave, $\vec{k}$ the wave vector and $\omega$ the angular frequency. In this form the classical field equations describe the wave properties of radiation but not pointlike excitations of the field. 

Consider plane waves propagating along the $x_3$-axis. In order to obtain pointlike solutions to the classical field equations, the next step is therefore to further constrain $A_\mu (x_\nu)$ through the C-R equations 
\begin{equation} \label{eq: cr5}
\frac{\partial A_\mu}{\partial x_1} \pm \imath \frac{\partial A_\mu}{\partial x_2}  = 0
\end{equation}
If the total constrained solution $A_\mu^c$ is assumed to be the sum of positive-energy $A_{\mu}^{\beta +}$ and negative-energy $A_{\mu}^{\beta -}$ fields belonging to each of the two $\mathbf{C}_{\beta}$ domains, this gives the decomposition
\begin{equation} \label{eq: em_points} 
A_{\mu}^c = A_{\mu}^{++} + A_{\mu}^{-+} + A_{\mu}^{+-} + A_{\mu}^{--} 
\end{equation}
Hence, each of these fields can have the product forms
\begin{equation} \label{eq: em_points_1} 
A_{\mu}^{\beta +}(x_3, t, z_\beta^k) = \int \frac{d^3k}{(2\pi)^{3/2} \sqrt{2\omega_k}} [a_{\beta}(\vec{k})\Phi_{\beta k}^{r}\epsilon_{\mu \beta}^{k} e^{\imath(\vec{k}\vec{x}-\omega t)}]
\end{equation}
\begin{equation} \label{eq: em_points_2} 
A_{\mu}^{\beta -}(x_3, t, z_\beta^{k*}) = \int \frac{d^3k}{(2\pi)^{3/2} \sqrt{2\omega_k}} [a_{\beta }^{*}(\vec{k})\Phi_{\beta k}^{s*}\epsilon_{\mu \beta}^{k*} e^{-\imath(\vec{k}\vec{x}-\omega t)}]
\end{equation}
where $\Phi_{\beta k}^{n}$ is defined in eq. (\ref{eq: total_symmetric}). Here, $A_{\mu}^{\beta +}$ and $A_{\mu}^{\beta -}$ are independent solutions of the classical field equations admitting the possibility the index $r$ is different from the index $s$.

Inserting eqs. (\ref{eq: em_points_1}) and  (\ref{eq: em_points_2}) into the Lorenz condition (\ref{eq: lorenz_gauge}) gives
\begin{equation} \label{eq: right_circular} 
\epsilon_{\mu \beta}^{k} = \frac{1}{\sqrt{2}}
\left( \begin{array}{c}
1 \\
\beta \imath \\
0 \\
0
\end{array} \right)
\end{equation}
showing electromagnetic waves confined to the $\mathbf{C_{+1}}$ and $\mathbf{C_{-1}}$ domains have left and right circular polarizations respectively. All other polarization states can be expressed as a linear combination of the two circular polarization states.

In QFT, creation $\hat{a}^\dagger$ and annihilation $\hat{a}$ operators act on a particle state $|\vec{k}, \beta, n \rangle$ to create and destroy photons such that $n$ is the number of photons having the same wave vector $\vec{k}$ and polarization state $\beta$. The core argument of this paper is the abstract particle state $|\vec{k}, \beta, n \rangle$ can be replaced using the explicit function $\Phi_{\beta k}^n(z_\beta)$ defined in eq. (\ref{eq: total_symmetric}). This gives
\begin{equation} \label{eq: vacuum}
|\vec{k}, \beta, n \rangle \rightarrow |z_\beta, \vec{k}, n \rangle = \Phi_{\beta k}^n(z_\beta)
\end{equation}
such that
\begin{equation} \label{eq: ladder_ops1}
\hat{a}\Phi_{\beta k}^n = \sqrt{n}\Phi_{\beta k}^{n-1}, \quad \hat{a}^\dagger\Phi_{\beta k}^n = \sqrt{n+1}\Phi_{\beta k}^{n+1}
\end{equation}
It is readily shown from eqs. (\ref{eq: ladder_ops1}) that
\begin{equation} \label{eq: ladder_ops2}
\hat{a}^\dagger \hat{a}\Phi_{\beta k}^n = n \Phi_{\beta k}^n
\end{equation}
where $\hat{N} = \hat{a}^\dagger \hat{a}$ is the occupancy number operator for the particle state.

The foregoing results can now be used to elevate the electromagnetic field (\ref{eq: em_points}) in to the form of the field operator
\begin{equation} \label{eq: em_quantum_field_opr} 
\hat{A}_{\mu}^c = \int \frac{d^3k}{(2\pi)^{3/2} \sqrt{2\omega_k}} \sum_{\beta} [\hat{a}^\dagger\epsilon_{\mu \beta}^{k} e^{\imath(\vec{k}.\vec{x}-\omega t)} + \hat{a}\epsilon_{\mu \beta}^{k*} e^{-\imath(\vec{k}.\vec{x}-\omega t)}]
\end{equation}
where
\begin{equation} \label{eq: coeff_1}
a_{\beta}(\vec{k})\Phi_{\beta k}^{r} = \hat{a}^\dagger |z_\beta, \vec{k}, n \rangle
\end{equation}
\begin{equation} \label{eq: coeff_2}
a_{\beta}^*(\vec{k})\Phi_{\beta k}^{s *} = \hat{a} |z_\beta, \vec{k}, n \rangle
\end{equation}
such that
\begin{equation} \label{eq: em_quantum_field}
A_{\mu}^c = \hat{A}_{\mu}^c |z_\beta, \vec{k}, n \rangle
\end{equation}
Eq. (\ref{eq: em_quantum_field_opr}) is a standard form for the electromagnetic quantum field operator. Eq. (\ref{eq: em_quantum_field}) is a solution to Maxwell's equations and Lorenz gauge condition including the field operator in it.

\section{Spin Angular Momentum}
The derivation for the quantized electromagnetic field above has furnished an explicit expression for the particle states. Eq. (\ref{eq: em_quantum_field}) equates the quantum field to a solution of the classical electromagnetic field equations indicating the quantum field is really just a disguised many-particle wavefunction. It is of interest therefore to investigate if the properties of the quantum field can be calculated using quantum mechanical operators applied to this many-particle wavefunction form of the quantum field.

It is a convenient starting point to write eq. (\ref{eq: em_quantum_field}) in the form
\begin{equation} \label{eq: em_quantum_field_compact} 
A_{\mu}^c = \int \frac{d^3k}{(2\pi)^{3/2} \sqrt{2\omega_k}} \sum_{\beta} \hat{A}_{\mu}^{\beta}(\vec{k})|z_\beta, \vec{k}, n \rangle
\end{equation}
where
\begin{eqnarray} \label{eq: comp_op}
\hat{A}_{\mu \beta}^{k}  = e^{+\imath (\vec{k}\vec{x}- \omega t)} \frac{1}{\sqrt{2}} \epsilon_{\mu}(\vec{k},\beta) \hat{a}^\dagger + e^{- \imath (\vec{k}\vec{x}- \omega t)} \frac{1}{\sqrt{2}} \epsilon_{\mu}^*(\vec{k},\beta) \hat{a}
\end{eqnarray} 
is the component electromagnetic field operator for a single polarization state $\beta$ and wave vector $\vec{k}$.

The quantum mechanical operators for momentum $\hat{p_3}$, energy $\hat{E}$ and angular momentum $\hat{L_3}$ of a single particle wavefunction take the form
\begin{equation} \label{eq: qm_ops1}
\hat{p}_3 = -\imath \hbar \frac{\partial}{\partial x_3}, \quad \hat{E} = \imath \hbar \frac{\partial}{\partial t}
\end{equation}
\begin{equation} \label{eq: ang_mom_opr}
\hat{L_3} =  \frac{\hbar}{\imath}\left( x_1 \frac{\partial}{\partial x_2} - x_2 \frac{\partial}{\partial x_1} \right)
\end{equation}
where $\hbar$ is Planck's constant divided by $2\pi$. It is significant that the translational energy-momentum of photons is contained in the electromagnetic field operator but the intrinsic spin is encoded in the particle state. The upshot is for more than one identical particle the energy-momentum operators (\ref{eq: qm_ops1}) must be modified to take account of the occupancy number of the particle state giving
\begin{equation} \label{eq: qm_ops2}
\hat{p}_3 = -\imath \hbar \hat{N}\frac{\partial}{\partial x_3}, \quad \hat{E} = \imath \hbar \hat{N} \frac{\partial}{\partial t}
\end{equation}
whereas no modification is needed to the orbital angular momentum operator. 

Eqs. (\ref{eq: qm_ops1}) and (\ref{eq: ang_mom_opr}) can be applied to an n-particle state to give momentum
\begin{equation} \label{eq: momentum1}
\langle p_3 \rangle = \int_{-z_0}^{+z_0} \int_{-z_0^*}^{+z_0^*} dz dz^* \langle z_\beta^*, \vec{k}, n |\hat{A}_{\mu \beta}^{k *} \hat{p}_3 \hat{A}_{\mu \beta}^{k}|z_\beta, \vec{k}, n \rangle = n \hbar k, 
\end{equation}
total energy
\begin{equation} \label{eq: energy1}
\langle E \rangle = \int_{-z_0}^{+z_0} \int_{-z_0^*}^{+z_0^*} dz dz^* \langle z_\beta^*, \vec{k}, n |\hat{A}_{\mu \beta}^{k *}  \hat{E} \hat{A}_{\mu \beta}^{k}|z_\beta, \vec{k}, n \rangle = n \hbar \omega,
\end{equation}
and spin angular momentum
\begin{equation} \label{eq: ang_mom1}
\langle s_3 \rangle = \int_{-z_0}^{+z_0} \int_{-z_0^*}^{+z_0^*} dz dz^* \langle z_\beta^*, \vec{k}, n |\hat{A}_{\mu \beta}^{k *}  \hat{L_3} \hat{A}_{\mu \beta}^{k}|z_\beta, \vec{k}, n  \rangle = n \beta \hbar
\end{equation}
having set $\kappa = 1$ to fix agreement between the predicted and experimental values for photon spin. The orbital angular momentum of a complex point function shall be interpreted as spin angular momentum for two reasons. One is that it originates from a point source. The other is the appearance of $\beta$ in eq. (\ref{eq: ang_mom1}) correlates the magnitude of the angular momentum to the circular polarization state of the photon. In particular, it is clear that complex point functions can only satisfy the Lorenz gauge condition (\ref{eq: lorenz_gauge}) for left or right circular polarization states. It is also clear these two solutions have opposite angular momentum calculated using the operator (\ref{eq: ang_mom_opr}).

\section{Concluding Remarks}
Complex point functions have been defined to be products of a real Gaussian function and a complex Dirac delta function. It has further been shown that the quantized electromagnetic field can be obtained as a solution of Maxwell's equations and Lorenz gauge condition by including complex point functions in it to represent the particle states. Particle states are constructed to be symmetric under the exchange of identical particles. Creation and annihilation operators assume their usual role in this formulation to add and subtract photons from the particle states.

An interesting feature of having explicit expressions for particle states is that quantum mechanical operators can be applied to them through the electromagnetic field operator to calculate properties of the quantum field. The orbital angular momentum of the complex point functions has been interpreted as spin angular momentum since it originates from a point source and correlates to the polarization state of the complex point function. Complex points functions only satisfy the Lorenz gauge condition for circular polarization states.

\end{document}